\documentclass[aps,prb,twocolumn,longbibliography,groupedaddress]{revtex4-1}

\usepackage{hyperref}
\usepackage{bbm}

\begin{document}

\title{Thermodynamic perturbation theory for non-interacting quantum particles with application
  to spin-spin interactions in solids}
\date{2018-02-08}
\author{Cezary \'{S}liwa}\email{sliwa@ifpan.edu.pl}
\affiliation{Institute of Physics, Polish Academy of Sciences,
Aleja Lotnik\'{o}w 32/46, PL-02668 Warsaw, Poland}

\author{Tomasz Dietl}
\affiliation{International Research Centre MagTop, Aleja Lotnik\'{o}w 32/46, PL-02668 Warsaw, Poland}
\affiliation{Institute of Physics, Polish Academy of Sciences,
Aleja Lotnik\'{o}w 32/46, PL-02668 Warsaw, Poland}
\affiliation{WPI-Advanced Institute for Materials Research, Tohoku University, Sendai 980-8577, Japan}

\begin{abstract}
The determination of the Landau free energy (the grand thermodynamic potential) by a perturbation theory
is advanced to arbitrary order for the specific case of non-interacting fermionic systems
perturbed by a one-particle potential. Peculiar features of the formalism are highlighted,
and its applicability for bosons is indicated.
The results are employed to develop a more explicit approach describing exchange interactions
between spins of Anderson's magnetic impurities in metals, semiconductors, and insulators. Within the fourth order
our theory provides on the equal footing formulae for
the Ruderman-Kittel-Kasuya-Yosida, Bloembergen-Rowland, superexchange, and two-electron exchange integrals
at non-zero temperature.
\end{abstract}

\maketitle

\section{Introduction}

Significant effort has been devoted since the 1950s to the development of a quantum perturbation
theory of thermodynamic potentials for gases of interacting fermions. The theory,
as presented in Refs.\,\onlinecite{Glassgold:1959_PR,Thouless:1959_PR,Kohn:1960_PR,Luttinger:1960_PR},
provides exact formulae for diagrammatic expansion of thermodynamic potentials
in powers of a two-particle interaction potential. The starting point of such a theory
is one-particle Hamiltonian, whose eigenstates
and eigenergies are known, for example, having been obtained by exact diagonalization
of the one-particle Hamiltonian.

In general, however, the diagonal form of one-particle Hamiltonian is unknown, and
one has to resort to some perturbation theory
to (approximately) diagonalize the Hamiltonian even in non-interacting cases.
The standard tool to perform such calculations, the Rayleigh-Schr\"{o}dinger (RS) quantum perturbation
theory, requires significant amounts of algebra at each consecutive order of the expansion.
Moreover, in practice, application of the standard RS perturbation theory may be complicated by possible
degeneracy in the set of zeroth-order eigenstates, which requires prescribing a gauge.

As we show in this Article, such a procedure is not necessary within the quantum perturbation theory applied
to the Landau free energy.  The equations we obtain for successive
orders of expansion in powers of the one-particle potential are relatively simple, and have a surprising feature which
manifests beyond the second order.

We demonstrate how this new formalism can serve to study exchange interactions in solids containing magnetic impurities
described by the Anderson Hamiltonian. In particular,
we provide  general formulae for exchange integrals between two Anderson's magnetic
impurities at non-zero temperature and show that in the leading (fourth) order they describe on the equal footing the
superexchange, two-electron, and electron-hole (Bloembergen-Rowland) mechanisms in insulators
and, additionally, the Ruderman-Kittel-Kasuya-Yosida coupling in metals and extrinsic semiconductors.

Our paper is organized as follows. We first recall, in Sec.\,II, the standard 4th order formula for the
energy and discuss its shortcomings. Our alternative approach is discussed qualitatively
in Sec.\,III, whereas the formal perturbation theory for the
Landau free energy is presented in Sec.\,IV. Its application for the case of exchange interactions
in solids containing Anderson's magnetic impurities is exposed in Sec.\,V.

\section{Standard approach}

The standard theory of exchange interactions between magnetic impurity spins in solids, as developed by Larson {\em et al.} \cite{Larson:PRB_1988} and Savoyant {\em et al.} \cite{Savoyant:2014_PRB},
following the work of Anderson, Falicov and others \cite{Anderson:1950_PR,Goodenough:1958_JPCS,Falicov:1993}, is based on the following formulation of the fourth-order quantum perturbation
theory for fermions: the matrix element of an effective Hamiltonian $H_{\text{eff}}$ between the initial state $\left| i \right>$ and the final state
$\left| f \right>$, for a particle subject to a perturbing interaction described by a one-particle operator $V$,
is given by the sum of paths over intermediate states $I_1$, $I_2$, $I_3$ (see Eq.\,4.3 of Ref.\,\onlinecite{Larson:PRB_1988}),
\begin{equation}
  \left< f \middle| H_{\text{eff}} \middle| i \right> = \sum_{I_1, I_2, I_3} \frac{\left< f \middle| V \middle| I_1 \right>
  \left< I_1 \middle| V \middle| I_2 \right> \left< I_2 \middle| V \middle| I_3 \right>
  \left< I_3 \middle| V \middle| i \right>}{(E_0 - E_1) (E_0 - E_2) (E_0 - E_3)}.
  \label{eq:Larson}
\end{equation}
Now, to calculate the exchange interaction energy, one assumes as $i$ the state with the first spin up and the second spin down, and as $f$ the state with the first spin down and the second spin up. This approach is valid at zero temperature and breaks down at energy crossings, $E_0 = E_i$. Moreover, it has been noted \cite{Gang:1993_JPA,Psiachos:2015_AP}
that taking in Eq.\,\ref{eq:Larson} the unperturbed ground state energy as $E_0$ is (in general) an inappropriate approximation.

\section{Alternative approach}

Here we develop an alternative  more formal and strict approach in which the Landau free energy is determined by the perturbation theory \cite{Landau:2001_B}.
Our approach is valid at non-zero temperature (as long as the perturbation is smaller than $k_B T$) and handles properly divergences associated with energy crossings, $E_0 = E_i$ in denominators
of Eq.\,(\ref{eq:Larson}).

Our model system
consists of two magnetic impurities with singly occupied $d$ orbitals hybridizing with extended band states
\emph{via} the hybridization operator $V = V_{\text{hyb}}$. We assume the local spins are in spin coherent states\cite{Radcliffe:1971_JPA}, parameterized by the spin direction, and calculate
the free energy of the electronic subsystem (the occupied band states) as the function of the directions of the localized spins. We need to calculate
the fourth order perturbation to the energy of a band state $I_0$
\begin{equation}
  \sum_{I_1, I_2, I_3} \frac{\left< I_0 \middle| V \middle| I_1 \right>
  \left< I_1 \middle| V \middle| I_2 \right> \left< I_2 \middle| V \middle| I_3 \right>
  \left< I_3 \middle| V \middle| I_0 \right>}{(E_0 - E_1) (E_0 - E_2) (E_0 - E_3)},
\end{equation}
and to sum over all the band states with the Fermi-Dirac occupation function factor $f(E)$ as follows
\begin{eqnarray}
  \lefteqn{E^{(4)} = \sum_{I_0, I_1, I_2, I_3} \frac{f(E_0)}{(E_0 - E_1) (E_0 - E_2) (E_0 - E_3)}  {}} \nonumber \\
   & & \qquad\qquad \times\left< I_0 \middle| V \middle| I_1 \right>
  \left< I_1 \middle| V \middle| I_2 \right> \left< I_2 \middle| V \middle| I_3 \right>
  \left< I_3 \middle| V \middle| I_0 \right>,
\end{eqnarray}
where now the final state coincides with the initial one, and all energies are the unperturbed ones.
We have verified directly using a computer algebra system that
the last equation is rigorous, with the requirement that the sum over the states $(I_0, I_1, I_2, I_3)$ has to be extended
over \emph{all} states, also including those combinations of $(I_0, I_1, I_2, I_3)$ for which the denominator vanishes, either because of the degeneracy of the unperturbed Hamiltonian or (more importantly) due to repetitions in the sequence $(I_0, I_1, I_2, I_3)$.
Indeed, although only $f(E_0)$ appears explicitly in the equation,
if l'H\^{o}pital's rule is applied in a straight-forward manner \cite{Werpachowska:2010_PRB},
the derivatives $f'(E_0)$,
$f''(E_0)$ and $f'''(E_0)$ appear, as required by the chain rule for the derivatives of $f(E(\lambda))$.
We detail the regularization prescription below.

\section{Thermodynamic perturbation theory for systems of non-interacting particles}

Consider now a system of $n_f$ non-interacting fermions numbered $n = 1, 2, \ldots, n_f$, with
the corresponding annihilation and creation operators $({\hat a}_n, {\hat a}_n^{\dagger})$
and eigenenergies $E_n^{(0)}$, subject to a one-particle perturbation
$\hat V = \sum_{mn} V_{mn} \, {\hat a}_m^{\dagger} {\hat a}_n$.
The Hamiltonian of the system reads
\begin{equation}
  \hat H(\lambda) = \sum_n E_n^{(0)} {\hat a}_n^{\dagger} {\hat a}_n + \lambda \hat V.
\end{equation}
The grand thermodynamic potential, called also the Landau free energy, assumes the form
\begin{eqnarray}
  \lefteqn{\Omega(\lambda) = - k_B T \, \mathop{\mathrm{Tr}} \Biggl(
    \ln \Biggl\{ \hat{\mathbbm{1}}  {}} \nonumber\\
    & & \qquad\qquad +\exp\left[-\frac{1}{k_B T} \left(\hat H_{1} (\lambda) - \mu \hat N_{1} \right)\right] \Biggr\} \Biggr),
\end{eqnarray}
where the lower subscript $1$ indicates a one-particle operator, $\mu$ is the chemical potential,
$\hat N = \sum_n {\hat a}_n^{\dagger} {\hat a}_n$ is the number of particles operator
($\hat N_{1}$ is the identity operator). We expand $\Omega(\lambda)$ in powers of $\lambda$,
\begin{equation}
  \Omega(\lambda) = \Omega^{(0)} + \lambda \, \Omega^{(1)} + \lambda^2 \, \Omega^{(2)} \ldots,
\end{equation}
where
\begin{eqnarray}
  \Omega^{(0)} & = & -k_B T \sum_n \ln \left\{ 1 + \exp\left[-\frac{1}{k_B T}\left(E_n^{(0)} - \mu \right)\right] \right\}, \quad \\
  \Omega^{(1)} & = & \sum_{n_1} f\left(E_{n_1}^{(0)}\right) V_{n_1 n_1} \label{eq: p1}, \\
  \Omega^{(2)} & = & \sum_{n_1, n_2}  \frac{f\left(E_{n_1}^{(0)}\right)}{E_{n_1}^{(0)} - E_{n_2}^{(0)}} V_{n_1 n_2} V_{n_2 n_1}, \label{eq: p2} \\
  \Omega^{(3)} & = & \sum_{n_1, n_2, n_3}  \frac{f\left(E_{n_1}^{(0)}\right)}{\left(E_{n_1}^{(0)} - E_{n_2}^{(0)}\right) \left(E_{n_1}^{(0)} - E_{n_3}^{(0)}\right)} \nonumber \\ & & \qquad \qquad {} \times V_{n_1 n_2} V_{n_2 n_3} V_{n_3 n_1},
  \label{eq: p3} \\
  \Omega^{(4)} & = & \sum_{n_1, n_2, n_3, n_4} \nonumber \\
   & & \frac{f\left(E_{n_1}^{(0)}\right)}{\left(E_{n_1}^{(0)} - E_{n_2}^{(0)}\right) \left(E_{n_1}^{(0)} - E_{n_3}^{(0)}\right) \left(E_{n_1}^{(0)} - E_{n_4}^{(0)}\right)}
  \nonumber \\ & & \qquad \qquad {} \times
  V_{n_1 n_2} V_{n_2 n_3} V_{n_3 n_4} V_{n_4 n_1}, \label{eq: p4} 
\end{eqnarray}
with the Fermi-Dirac distribution function
\begin{equation}
  f(E) = \frac{1}{1 + \exp\left(\frac{E - \mu}{k_B T}\right)}.
\end{equation}
This result is partly formal, because the denominators may vanish, \emph{e.g.} they do if there are repetitions in the sequence of summation indices.
We will explain the exact meaning of these expressions in such cases below, and give a more explicit formula in terms of a ratio of simple determinants
in Appendix \ref{app: recursion}.

By symmetrizing with respect to the summation indices we obtain the same expression in a more standard notation,
\begin{equation}
  \Omega^{(2)} = \frac{1}{2} \sum_{m, n}  \frac{f\left(E_{m}^{(0)}\right) - f\left(E_{n}^{(0)}\right)}{E_{m}^{(0)} - E_{n}^{(0)}} \left| V_{mn} \right|^2,
  \label{eq: o2symm}
\end{equation}
sometimes also written as\cite{Lewiner:1980_JPC}
\begin{equation}
  \Omega^{(2)} = \sum_{m, n}  \frac{f\left(E_{m}^{(0)}\right) \left[ 1 - f\left(E_{n}^{(0)}\right) \right]}{E_{m}^{(0)} - E_{n}^{(0)}} \left| V_{mn} \right|^2. \label{eq: 1minus}
\end{equation}
However, surprisingly, the higher orders [Eqs.\,(\ref{eq: p3}) and (\ref{eq: p4})] lack the ``expected'' factors
of $\prod_{i > 1} \left[ 1 - f\left(E_{n_i}^{(0)}\right) \right]$ [cf.\,Eq.\,(\ref{eq: 3bad})],
with $i$ corresponding to intermediate states other than the state being perturbed.
We interpret this fact as follows: the expectation behind writing $\Omega^{(2)}$ as in Eq.\,(\ref{eq: 1minus}) involves the Pauli principle, which would require the intermediate states to be empty to allow virtual transitions to those states. However, since the eigenstates of the full Hamiltonian remain orthogonal on perturbation, an admixture of an occupied eigenstate
(which is just other wording for a virtual transition to that state) does \emph{not} violate the Pauli principle.
Moreover, \emph{e.g.} $\Omega^{(3)}$ given in Eq.\,(\ref{eq: p3}) obeys a particle-hole symmetry, not shared by its modification [Eq.\,(\ref{eq: 3bad})].

Most importantly, however, various literature theoretical prescriptions
(see, \emph{e.g.}, Lewiner, Gaj, and Bastard \cite{Lewiner:1980_JPC}) do not specify a procedure to handle singular
denominators, like those which occur identically for $m = n$ [$(\nu, k) = (\nu', k')$ in Lewiner's {\em et al.} Eq.\,2].
Our observation is that properly regularizing and including such terms in the total r.h.s. is crucial
to be able to interpret the result as a perturbation to the free energy of the system. Indeed, the latter includes --- via the chain rule --- the derivatives $f'(E_n)$, which are properly accounted for by applying
l'H\^{o}pital's rule to the terms in Eq.\,(\ref{eq: o2symm})  with $m = n$.

We stress that the above formulae are valid for any degeneracy in the set of energies. However, since the denominators are singular if degeneracies are present, the expressions may need a regularization. For example
\begin{eqnarray}
  \lefteqn{E_1, E_2 \to E \Rightarrow \frac{f\left(E_1\right)}{E_1 - E_2} + \frac{f\left(E_2\right)}{E_2 - E_1}  {}} \nonumber
   \\ & & \qquad\qquad =\frac{f\left(E_1\right) - f\left(E_2\right)}{E_1 - E_2} \to f'(E).
\end{eqnarray}

At any order, prior to regularization the iterated sums in Eqs.\,(\ref{eq: p2})
to (\ref{eq: p4}) have to be \emph{symmetrized} with respect to summation indices.
That is, one calculates the total of all terms corresponding to combinations $(n_1, n_2, \ldots)$
which differ by a permutation. Any permutation is allowed. Indeed, the fraction consisting
of the Fermi-Dirac numerator and the energetic denominators is preserved by any permutation which
keeps the first index in place, while the remaining product of matrix elements is preserved
by a cycle of all the indices (and these two kinds of permutations generate the full symmetric group).
Below we show that despite singular denominators in the individual terms,
each total is a well defined expression. In the following we assume that $x_1, x_2, \ldots \to x$,
$y_1, y_2, \ldots \to y$, $z_1, z_2, \ldots \to z$.
Possible combinations of the degeneracy in the set of up to 4 energies are summarized in Table \ref{tab: deg},
which serves as an index to the relevant equations.

\begin{widetext}

\begin{equation}
  \frac{f(x_1)}{x_1 - x_2} + \frac{f(x_2)}{x_2 - x_1} \to f'(x), 
  \label{eq: reg xx}
\end{equation}
\begin{eqnarray}
  \lefteqn{\frac{f(y_1)}{(y_1-x_1)(y_1-x_2)} + \frac{f(x_1)}{(x_1-y_1)(x_1-x_2)}  {}}    \nonumber \\
   & & \qquad \qquad \qquad \qquad {} + \frac{f(x_2)}{(x_2-y_1)(x_2-x_1)} \to
  \frac{\frac{f(x)-f(y)}{x-y}-f'(x)}{y-x},
\label{eq: reg xxy} \end{eqnarray}
\begin{eqnarray}
  \lefteqn{\frac{f(x_1)}{(x_1-x_2)(x_1-x_3)} + \frac{f(x_2)}{(x_2-x_1)(x_2-x_3)} + {}}\nonumber \\
  & & \qquad \qquad \qquad \qquad {} + \frac{f(x_3)}{(x_3-x_1)(x_3-x_2)} \to
  \frac{f''(x)}{2}, \label{eq: reg xxx}
\end{eqnarray}
\begin{eqnarray}
  \lefteqn{\frac{f(x_1)}{(x_1-y_1)(x_1-x_2)(x_1-z_1)} + \frac{f(y_1)}{(y_1-x_1)(y_1-x_2)(y_1-z_1)} + {}}
  \nonumber  \\
   \lefteqn{\qquad \qquad {} + \frac{f(x_2)}{(x_2-x_1)(x_2-y_1)(x_2-z_1)} + \frac{f(z_1)}{(z_1-x_1)(z_1-y_1)(z_1-x_2)} \to {}} \nonumber \\ & &
   \frac{1}{2} \frac{f(y) - f(z)}{y - z} \left[ \frac{1}{(y-x)^2} + \frac{1}{(z-x)^2} \right]
    + \frac{1}{(y-x)(z-x)} \Biggl[ f'(x) + {} \Biggr. \nonumber \\ & & \qquad \qquad \qquad \qquad \qquad \qquad
    \Biggl. {} + \left( f(x) - \frac{f(y) + f(z)}{2} \right)\left( \frac{1}{y-x} + \frac{1}{z-x} \right) \Biggr],
\label{eq: reg xxyz}
\end{eqnarray}
\begin{eqnarray}
  \lefteqn{\frac{f(x_1)}{(x_1-x_2)(x_1-y_1)(x_1-y_2)} + \frac{f(x_2)}{(x_2-x_1)(x_2-y_1)(x_2-y_2)} + {}}  \nonumber \\
   \lefteqn{\qquad {} + \frac{f(y_1)}{(y_1-x_1)(y_1-x_2)(y_1-y_2)} + \frac{f(y_2)}{(y_2-x_1)(y_2-x_2)(y_2-y_1)} \to {}} \nonumber  \\ & & \qquad \qquad \qquad \qquad \qquad \qquad \qquad \qquad
   \quad \frac{[f'(x)+f'(y)] - 2 \frac{f(x)-f(y)}{x-y}}{(x-y)^2},  \label{eq: reg xxyy}
\end{eqnarray}
\begin{eqnarray}
  \lefteqn{\frac{f(x_1)}{(x_1-x_2)(x_1-x_3)(x_1-y_1)} + \frac{f(x_2)}{(x_2-x_1)(x_2-x_3)(x_2-y_1)} + {}}   \nonumber  \\
   \lefteqn{\qquad \qquad {} + \frac{f(x_3)}{(x_3-x_1)(x_3-x_2)(x_3-y_1)} + \frac{f(y_1)}{(y_1-x_1)(y_1-x_2)(y_1-x_3)} \to {}} \nonumber \\ & &
    \qquad \qquad \qquad \qquad \qquad \qquad \qquad \qquad \qquad \qquad \qquad
   \frac{\frac{\frac{f(x)-f(y)}{x-y}- f'(x)}{y-x} - \frac{f''(x)}{2}}{y-x},
   \label{eq: reg xxxy}
\end{eqnarray}
\begin{eqnarray}
  \lefteqn{\frac{f(x_1)}{(x_1-x_2)(x_1-x_3)(x_1-x_4)} + \frac{f(x_2)}{(x_2-x_1)(x_2-x_3)(x_2-x_4)} + {}}  \nonumber \\
   & & {} + \frac{f(x_3)}{(x_3-x_1)(x_3-x_2)(x_3-x_4)} + \frac{f(x_4)}{(x_4-x_1)(x_4-x_2)(x_4-x_3)} \to \frac{1}{6} f'''(x).
\label{eq: reg xxxx}
\end{eqnarray}

\end{widetext}

\begin{table}
\begin{tabular}{||c|c|c||}
\hline
order& degeneracy& equation number\\
\hline
2& $xx$& (\ref{eq: reg xx})\\
3& $xxy$& (\ref{eq: reg xxy})\\
3& $xxx$& (\ref{eq: reg xxx})\\
4& $xxyz$& (\ref{eq: reg xxyz})\\
4& $xxyy$& (\ref{eq: reg xxyy})\\
4& $xxxy$& (\ref{eq: reg xxxy})\\
4& $xxxx$& (\ref{eq: reg xxxx})\\
\hline
\end{tabular}
\caption{Summary of possible degeneracies in a set of up to four eigenenergies.}
\label{tab: deg}
\end{table}

To prove Eqs.\,(\ref{eq: p1})--(\ref{eq: p4}) we make use of the relation given in Appendix \ref{app: part} [Eq.\,(\ref{eq: fdet})],
\begin{equation}
  \frac{1}{2} \mathop{\mathrm{Tr}} (H_0 + \lambda V) + \Omega(\lambda) = -\frac{1}{\beta} \ln \det \left[ - \frac{\partial}{\partial \tau} - (H_0 + \lambda V) \right],
\end{equation}
where $H_0$ is the unperturbed Hamiltonian and $V$ is the perturbation
(as usual, $\beta = 1 / k_B T$ and we drop the constant factor $\det \left[ - \frac{\partial}{\partial \tau} \right]$ which affects only the normalization of the partition function). Let
\begin{equation}
  \mathcal{G}_0^{-1} = - \frac{\partial}{\partial \tau} - H_0, \qquad \delta \mathcal{G}^{-1} = -\lambda V.
\end{equation}
We rewrite the above relation as
\begin{equation}
   \frac{1}{2} \mathop{\mathrm{Tr}} (H_0 + \lambda V) + \Omega(\lambda) = -\frac{1}{\beta} \mathop{\mathrm{Tr}} \left\{ \ln \left[ \mathcal{G}_0^{-1} + \mathcal{G}^{-1} \right] \right\},
\end{equation}
which can be easily expanded into power series \cite{Konig:2001}:
\begin{eqnarray}
  \lefteqn{\frac{1}{2} \mathop{\mathrm{Tr}} (H_0 + \lambda V) + \Omega(\lambda)} \nonumber \\ & = & \left[ \frac{1}{2} \mathop{\mathrm{Tr}} (H_0) + \Omega(0) \right]
+ \mathop{\mathrm{Tr}} \left[ \frac{(-1)^{n}}{n} \left( \mathcal{G}_0 \mathcal{G}^{-1} \right)^n \right]. \label{eq: fdetham}
\end{eqnarray}
In the spectral representation,
\begin{eqnarray}
  \lefteqn{\mathop{\mathrm{Tr}} \left[ \left( \mathcal{G}_0 \mathcal{G}^{-1} \right)^n \right] =
    \sum_{m = -\infty}^{\infty} \sum_{i_1} \sum_{i_2} \ldots \sum_{i_n} } \nonumber\\
    & & (-\lambda)^n \left[ \prod_{a = 1}^n \frac{1}{i \omega_m - (E_{i_a} - \mu)} \right] V_{i_1 i_2} V_{i_2 i_3} \ldots V_{i_n i_1},
\end{eqnarray}
where $\omega_m = (2 m + 1) \pi / \beta$ is the fermionic Matsubara frequency, and $i_1, i_2, \ldots, i_n$ label eigenstates of the unperturbed Hamiltonian. The quantity (double braces indicate a multiset)
\begin{eqnarray}
  \lefteqn{s_n(\left\{\left\{ E_{i_1}, E_{i_2}, \ldots, E_{i_n} \right\}\right\})}  \nonumber\\
  & = &  \frac{1}{\beta} \sum_{m = -\infty}^{\infty} \prod_{a = 1}^n \frac{1}{i \omega_m - (E_{i_a} - \mu)},
\label{eq: matsum}
\end{eqnarray}
can be calculated in the standard manner as a contour integral \cite{Nieto:1995_CPC} over $z = i \omega_m$, where the integrand is the summand of Eq.\,(\ref{eq: matsum}) multiplied by the occupation function $f(z)$. Assuming no degeneracy, each residue at $E_{i_a} \in \left\{\left\{ E_{i_1}, E_{i_2}, \ldots, E_{i_n} \right\}\right\}$ contributes to the value of $s_n$ [Eq.\,(\ref{eq: matsum})] the term
\begin{equation}
  f(E_{i_a}) \mathop{\prod\limits_{b = 1}^n}\limits_{b \ne a} \frac{1}{E_{i_a} - E_{i_b}},
\end{equation}
[cf.\,Eqs.\,(\ref{eq: p1})--(\ref{eq: p4})],
while for $n = 1$ the contour at infinity yields a contribution of $1/2$ which cancels the term $\frac{1}{2} \mathop{\mathrm{Tr}} (\lambda V)$ on the l.h.s. of Eq.\,(\ref{eq: fdetham}).

If the set of energies is degenerate, \emph{e.g.} $E_{i_{a}} = E_{i_{b}}$, higher order singularities are present in the integrand. Either the residue at a higher order singularity can be obtained as a derivative of the regular part of the integrand at $z = E_{i_{a}} = E_{i_{b}}$, or the limit of $s_n$ can be taken as $E_{i_{b}} \to E_{i_{a}}$. If a given eigenstate of the unperturbed Hamiltonian appears multiple times, \emph{e.g.} $i_{a} = i_{b}$, then $E_{i_{a}} = E_{i_{b}}$, as if a degeneracy was present. In such a case one can either use
the former prescription, or take the limit of $s_n$ as $E_{i_{b}} \to E_{i_{a}}$, paying careful attention
to consider the symbols $E_{i_{a}}$ and $E_{i_{b}}$ in Eq.\,(\ref{eq: matsum})  as independent variables. This ends the proof.

Equations (\ref{eq: reg xx})--(\ref{eq: reg xxxx}) can be generalized to higher orders by
using the recursion relation given in Appendix \ref{app: recursion} [Eq.\,(\ref{eq: recurs})].

We have verified in the second order of perturbation
that our result is valid for bosons, with the Fermi-Dirac occupation function replaced by its bosonic counterpart.

A relation of our results
to the Rayleigh-Schr\"{o}dinger perturbation theory and to the path integral formulation
is discussed in Appendices \ref{app: RS} and \ref{app: path}, respectively.

\section{Spin interactions in solids}

Let us now return to the specific case of a spin-spin interaction in a solid containing magnetic impurities described by the Anderson Hamiltonian. The Hilbert space of our model system is a direct sum of the sector of localized electronic states (for example $3d$ states of a magnetic dopant) and the continuum (band states). The localized states are spin-degenerate. This describes the situation where a magnetic impurity (such as Mn) is present in a solid matrix (specifically a semiconductor such as CdTe or HgTe). Hence, the Hamiltonian (the Anderson Hamiltonian for multiple impurities),
\begin{equation}
  H = \sum_k E_k \, a_{k}^{\dagger} a_{k} + \sum_l H_l + \lambda ( V_{hyb} + V_{\text{hyb}}^{\dagger} ),
\end{equation}
where $\{ E_k \}$ is the continuous spectrum, and the Hamiltonian of an impurity with label $l \in L$, $H_l$, is of the form
\begin{equation}
H_l = E_d \, ( a_{l\uparrow}^{\dagger} a_{l\uparrow} + a_{l\downarrow}^{\dagger} a_{l\downarrow} ) + U \, a_{l\uparrow}^{\dagger} a_{l\uparrow} a_{l\downarrow}^{\dagger} a_{l\downarrow}.
\end{equation}
The one-particle operator $V_{\text{hyb}}$ describes virtual transitions
from the extended to the localized states. The term $V_{\text{hyb}} + V_{\text{hyb}}^{\dagger}$
is considered as a perturbation to the electronic Hamiltonian. Hence, our perturbation theory formalism applies.
One easily observes that a process corresponding to a given set of states $(n_1, n_2, \ldots)$ can contribute to a free energy perturbation if the states $(n_1, n_2, \ldots)$ are alternately extended and localized,
and that the leading order in which a spin-spin interaction occurs is the fourth one. Moreover,
in the leading contributing processes, the labels $(n_1, n_2, n_3, n_4)$ correspond to two $d$ states
$(d_1, d_2)$ associated with two different impurities and two band states $k$ and $k'$,
thus the only possible repetition is $k = k'$. (This observation stays behind the success
of the standard approach to spin-spin interactions in solids
and demonstrates that the standard approach may fail at higher orders).
We denote by $E_k$ and $E_{k'}$ the energies of the band states, by $E_d$ the energy
of the singly-occupied localized $d$ state, and by $U$ the Coulomb energy of a doubly-occupied $d$
state (this energy adds to $E_d)$. Our goal is calculate the exchange integral describing the interaction
between two localized spins. Therefore, we assume the localized states are occupied by electrons
with spin either up or down, and consider the spins of those electrons as the interacting objects, and the remaining electrons as a non-interacting gas. Thus, the present approach neglects the potential exchange  within the carriers, which augments the ferromagnetic portion of RKKY coupling \cite{Dietl:1997_PRB}, as well as between carriers and localized spins, in particular the intra-ion $sp-d(f)$ exchange that may control the strength of spin-spin interactions between rare earth impurities \cite{Geertsma:1990_PB}, for which $V_{\text{hyb}}$ is small.

We calculate the free energy of the non-interacting electronic gas as the total of the free energy of the electronic states with spin up and those with spin down (this a further simplification which is valid if no spin-orbit interaction is present). If the directions of the local spins agree,
and the direction of the electronic spin is the same, we use Eq.\,(\ref{eq: reg xxyz}) with $x = E_d$ (the singly-occupied $d$ state --- double occupation is prohibited by the Pauli principle),
$y = E_k$ and $z = E_{k'}$ (the two intermediate band states). We do so, because since the dopants are identical, the $d$ states are degenerate.
If the spin of the electron is opposite to the direction of the local spins,
$x = E_d + U$ (the unoccupied $d$ state energy, including the Coulomb interaction with the electron in the $d$ state which is considered as the local spin). However, if the directions of the local spins are opposite, the degeneracy of the two intermediate states is lifted,
and Eq.\,(\ref{eq: p4}) can be used directly with the energies $E_k$, $E_{k'}$, $E_d$ and $E_d + U$. We calculate the exchange interaction energy as the difference
of the electronic free energies for agreeing and opposite directions of the local spins.
Thus the Fermi-Dirac occupation function and the energetic denominators contribute the factor (let us denote it $A_{k,k'}$):
\begin{widetext}
\begin{eqnarray}
  A_{k,k'} & = & \frac{1}{U} \left[ \frac{2 E_d^2 + 2 E_k E_{k'} + (E_k + E_{k'}) U - 2 E_d (E_k + E_{k'} +U)}{(E_d - E_k)^2 (E_d - E_{k'})^2} f\left( \frac{E_d - \mu}{k_B T} \right)  + {} \right. \nonumber\\ & & \left. {} + \frac{-2 (E_d - E_k) (E_d - E_{k'}) + 3 (E_k + E_{k'} - 2 E_d) U - 4 U^2}{(E_d + U - E_k)^2 (E_d + U - E_{k'})^2} f\left( \frac{E_d + U - \mu}{k_B T} \right) \right] + {} \nonumber \\ & & {} +
  U^2 \left[ \frac{1}{(E_d - E_k)^2 (E_d - E_k + U)^2} \frac{f\left( \frac{E_k - \mu}{k_B T} \right)}{E_k - E_{k'}} + \frac{1}{(E_d - E_{k'})^2 (E_d - E_{k'} + U)^2} \frac{f\left( \frac{E_{k'} - \mu}{k_B T} \right)}{E_{k'} - E_k} \right] + {} \nonumber \\
  & & {} + \frac{1}{k_B T} \left[ \frac{1}{(E_d - E_k) (E_d - E_{k'})} f'\left( \frac{E_d - \mu}{k_B T} \right) + \frac{1}{(E_d + U - E_k) (E_d + U - E_{k'})} f'\left( \frac{E_d + U - \mu}{k_B T} \right) \right].
\label{eq: aakkp}
\end{eqnarray}
\end{widetext}

The final expression for the difference between the free energy of the carriers for parallel local spins [$\Omega_c(\uparrow \uparrow)$]
and for antiparallel local spins [$\Omega_c(\uparrow \downarrow)$] is proportional to the matrix element of $V_{\text{hyb}}$ in the fourth power:
\begin{widetext}
\begin{eqnarray}
  \Omega_c(\uparrow \uparrow) - \Omega_c(\uparrow \downarrow) & = &
    \sum_{k,k'} A_{k,k'} \left< k \middle| V_{\text{hyb}}^{\dagger} \middle| d_1 \right> \left< d_1 \middle| V_{\text{hyb}} \middle| k' \right>
    \left< k' \middle| V_{\text{hyb}}^{\dagger} \middle| d_2 \right> \left< d_2 \middle| V_{\text{hyb}} \middle| k \right> \\
    & = & \frac{1}{2} \sum_{k,k'} A_{k,k'} \biggl(
    \left< k \uparrow \middle| V_{\text{hyb}}^{\dagger} \middle| d_1 \uparrow \right> \left< d_1 \uparrow\middle| V_{\text{hyb}} \middle| k' \uparrow \right>
    \left< k' \uparrow \middle| V_{\text{hyb}}^{\dagger} \middle| d_2 \uparrow\right> \left< d_2 \uparrow\middle| V_{\text{hyb}} \middle| k \uparrow \right> + {} \label{eq: fupdown1} \nonumber \\
    & & \qquad {} + \left< k \downarrow \middle| V_{\text{hyb}}^{\dagger} \middle| d_1 \downarrow \right> \left< d_1 \downarrow \middle| V_{\text{hyb}} \middle| k' \downarrow \right>
    \left< k' \downarrow \middle| V_{\text{hyb}}^{\dagger} \middle| d_2 \downarrow \right> \left< d_2 \downarrow \middle| V_{\text{hyb}} \middle| k \downarrow \right> \biggr).
\label{eq: fupdown}
\end{eqnarray}
\end{widetext}
We have used the fact that in the absence of a spin-orbit interaction the matrix elements factorize into the product of the orbital and the spin part,
\begin{equation}
  \left< d_i \uparrow \middle| V_{\text{hyb}} \middle| k \uparrow \right> = \left< d_i \downarrow \middle| V_{\text{hyb}} \middle| k \downarrow \right> =
  \left< d_i \middle| V_{\text{hyb}} \middle| k \right>,
\end{equation}
\begin{equation}
  \left< d_i \uparrow \middle| V_{\text{hyb}} \middle| k \downarrow \right> = \left< d_i \downarrow \middle| V_{\text{hyb}} \middle| k \uparrow \right> = 0,
\end{equation}
and only the orbital parts $\left< d_i \middle| V_{\text{hyb}} \middle| k \right>$ enter the equation for $\Omega_c(\uparrow \uparrow) - \Omega_c(\uparrow \downarrow)$.

Hence, the non-interacting approximation to the classical exchange integral between the two spins
is given by Eqs.\,(\ref{eq: aakkp}) and (\ref{eq: fupdown}) \emph{via}
\begin{equation}
  J_{12} \approx - \Bigl[ \Omega_c(\uparrow \uparrow) - \Omega_c(\uparrow \downarrow) \Bigr].
\end{equation}
Then, the spin-dependent part of the effective Hamiltonian for two spins can be reconstructed
from its spin coherent state\cite{Radcliffe:1971_JPA} representation as
\begin{equation}
  \hat{H}^{(4)}_{\text{eff}} = -2 J_{12} \, \hat{S}_1 \cdot \hat{S}_2,
\end{equation}
with spin $S = 1/2$ angular momentum operators $\hat{S}_1$ and $\hat{S}_2$ representing the degrees of freedom of the two spins.

In a crystalline solid, the matrix elements include the plane-wave phase factor:
\begin{equation}
  \left< d_i \middle| V_{\text{hyb}} \middle| k \right> = \exp(-i \kappa \cdot R_i) \tilde V_k,
\end{equation}
with $R_i$ being the real-space position of the impurity, and $\kappa$ the wave vector corresponding to the band state $k$. Therefore, the product of matrix elements is proportional to $\exp[-i (\kappa - \kappa') \cdot (R_2 - R_1)]$, and we obtain
\begin{eqnarray} \lefteqn{
  \Omega_c(\uparrow \uparrow) - \Omega_c(\uparrow \downarrow) = \sum_{k,k'} {} } \nonumber \\
  & & \quad {} \exp[-i(\kappa - \kappa') \cdot (R_2 - R_1)] \left| \tilde V_k \right|^2 \left| \tilde V_{k'} \right|^2 A_{k,k'}.
\label{eq: fupdown2}
\end{eqnarray}
Here, $\exp[-i(\kappa - \kappa') \cdot (R_2 - R_1)]$ can be replaced with $\cos[(\kappa - \kappa') \cdot (R_2 - R_1)]$
by the possibility of symmetrization with respect to the interchange $k \leftrightarrow k'$. In general, if a spin orbit interaction
is present, and the spin directions are not collinear, this symmetrization must be suppressed.

Additional degeneracies may be present. In particular, in the limit $E_{k'} \to E_k$ we obtain:
\begin{eqnarray}
  \lefteqn{A_{k,k} = \frac{2 U^2 (2 E_d - 2 E_k + U)}{(E_d - E_k)^3 (E_d + U - E_k)^3} f\left( \frac{E_k - \mu}{k_B T} \right) {}} \nonumber \\ & & {} + \frac{1}{k_B T} \frac{U^2}{(E_d - E_k)^2 (E_d + U - E_k)^2} f'\left( \frac{E_k - \mu}{k_B T} \right) \nonumber
  \\  & & {} + \ldots \label{eq: rkky}
\end{eqnarray}
The second term is proportional to $\frac{1}{k_B T} f'$, \emph{i.e.} to the Dirac delta at the Fermi level in the zero-temperature limit.
In insulators (no density of states at the Fermi level), its contribution vanishes as $T \to 0$. However, this is not the case in metals, where this term represents the Ruderman-Kittel-Kasuya-Yosida (RKKY) interaction \cite{Ruderman:1954_PR}.
In contrast, Eq.\,4.5 of Ref.\,\onlinecite{Larson:PRB_1988} fails even to produce a meaningful (finite)
result for $\varepsilon_n(k) = \varepsilon_{n'}(k')$ ($E_k = E_{k'}$).

Although the resulting expression for $A_{k,k'}$
is valid for arbitrary temperature, we take the zero temperature limit in order to single out various contributions
discussed in the literature. Furthermore, we assume the case of an insulator, which allows us to omit terms proportional to the derivatives of the Fermi-Dirac occupation function.

Following literature conventions, we decompose our zero-temperature exchange integrals into three terms, $J_{dd} = J_{hh} + J_{eh} + J_{ee}$.
The first term is called superexchange and corresponds to occupied intermediate band states. The last term, called the two-electron term,
corresponds to empty intermediate band states. The remaining term, called the electron-hole term, turns out in fact
to comprise two contributions ($eh$ and $he$), and describes the Bloembergen-Rowland interaction \cite{Bloembergen:1955_PR}. The total numerical factor resulting from the energy denominators is
\begin{eqnarray}
 A_{k,k'} & \to & \Theta(\mu - E_k) \Theta(\mu - E_{k'}) A^{hh} \nonumber \\
 & & {} + \Theta(\mu - E_k) \Theta(E_{k'} - \mu) A^{he} \nonumber \\
 & & {} + \Theta(E_k - \mu) \Theta(\mu - E_{k'}) A^{eh} \nonumber \\
 & & {} + \Theta(E_k - \mu) \Theta(E_{k'} - \mu) A^{ee}. \label{eq: akkp0}
\end{eqnarray}
where the individual terms are given below ($A^{he}$ corresponds to
an occupied state with energy $E_k$, while $A^{eh}$ to an occupied state with energy $E_{k'}$).
\begin{widetext}
\begin{equation}
  A^{hh} = \frac{2}{(E_d + U - E_k)(E_d + U - E_{k'}) U} + \frac{1}{(E_d + U - E_k)(E_d + U - E_{k'})} \left(
    \frac{1}{E_d + U - E_k} + \frac{1}{E_d + U - E_{k'}} \right), \label{eq: ahh}
\end{equation}
\begin{equation}
  A^{he} = \frac{2}{(E_d + U - E_k)(E_d - E_{k'}) U} + \frac{1}{E_k - E_{k'}} \left(
    \frac{1}{E_d - E_{k'}} - \frac{1}{E_d + U - E_{k}} \right)^2, \label{eq: ahe}
\end{equation}
\begin{equation}
  A^{eh} = \frac{2}{(E_d - E_k)(E_d + U - E_{k'}) U} + \frac{1}{E_{k'} - E_k} \left(
    \frac{1}{E_d - E_k} - \frac{1}{E_d + U - E_{k'}} \right)^2, \label{eq: aeh}
\end{equation}
\begin{equation}
  A^{ee} = \frac{2}{(E_d - E_k)(E_d - E_{k'}) U} - \frac{1}{(E_d - E_k)(E_d - E_{k'})} \left(
    \frac{1}{E_d - E_k} + \frac{1}{E_d - E_{k'}} \right). \label{eq: aee}
\end{equation}
\end{widetext}
We underline that Eqs.\,(\ref{eq: ahh})--(\ref{eq: akkp0}) do not take into account the RKKY term [see Eq.\,(\ref{eq: rkky})] included in Eq.\,(\ref{eq: aakkp}).

The series of Eqs.\,(\ref{eq: ahh})--(\ref{eq: aee}) lead to exchange integrals in agreement
with formulae obtained by Blinowski, Kacman, and Majewski for $d^5$ transition-metal ions in insulators \cite{Blinowski:1996_unp}.
This agreement reconfirms the fact that luckily the regularization procedure is not necessary in this case.
These formulae were later modified for the $d^4$ configuration \cite{Blinowski:1996_PRB}.
The theoretical model for $d^4$ ions
was successfully employed to describe ferromagnetic ordering of Mn spins observed experimentally
in insulating $(\mathrm{Ga},\mathrm{Mn})\mathrm{N}$ with $\mathrm{Mn}^{3+}$ ions \cite{Stefanowicz:2013_PRB,Simserides:2014_EPJ}.
This agreement between theoretical and experimental results indicates that inclusion of higher order terms
is not necessary, at least when the short range superexchange dominates, the case of $(\mathrm{Ga},\mathrm{Mn})\mathrm{N}$.

The key accomplishment of this section is presented in Eq.\,\ref{eq: aakkp} that allows describing
various contributions to the exchange coupling of magnetic impurity pairs on an equal footing and at arbitrary
temperature. However, in the case of insulators the low temperature approximation usually holds,
since the energy distance of the Fermi level to both band edges and $d$ states is significantly larger than $k_B T$.
A similar approximation is metals is valid (as long as the RKKY term is being kept) as the
condition of a strong degeneracy of the carrier gas is fulfilled, $\mu \gg k_B T$.
In contrast to metals, however, this condition is often not satisfied in extrinsic
semiconductors \cite{Dietl:2014_RMP}, and indeed non-standard temperature effects
in the latter have been noted \cite{Boukari:2002_PRL}.

\section{Conclusions}

Landau's thermodynamic perturbation theory \cite{Landau:2001_B}
is a simple-to-use, general, strict, formal, systematic and elegant tool.
Here, it has been advanced to arbitrary order in the specific case of non-interacting fermionic systems
(its extension to non-interacting bosonic systems appears as straightforward).
The developed approach has been applied to the case of exchange interactions
between Anderson's magnetic impurities in solids.
In the fourth order our results generalize literature formulae to non-zero temperatures and handle
properly zeros of the energetic denominators due to repeated level indices and possible
overlaps or crossings of energy levels. In particular,
the present results apply rigorously to situations, in which band carriers
mediate a long-ranged interaction between localized spins, the case of dilute magnetic metals and extrinsic
dilute magnetic semiconductors.

\section{Acknowledgments}

The work has been supported within the Master project of National Center of Science in Poland (2011/02/A/ST3/00125).
The International Centre for Interfacing Magnetism and Superconductivity with
Topological Matter project is carried out within the International Research Agendas
programme of the Foundation for Polish Science co-financed by the European Union
under the European Regional Development Fund.

\appendix

\section{Partition function as a functional determinant}
\label{app: part}

The partition function for fermions can be represented
as a functional determinant in the imaginary time formalism,
\begin{equation}
  1 + \exp(-\beta E) = \exp(-\beta E / 2) \frac{\det\left( -\frac{\partial}{\partial \tau} - E \right)}{\det\left( -\frac{\partial}{\partial \tau}\right)}.
\label{eq: fdet}
\end{equation}
Indeed, the ratio on the r.h.s. of this relation can be represented as follows and making use of the Euler representation for the meromorphic function $\sin(z)$ we obtain:
\begin{eqnarray}
  \lefteqn{\frac{\det\left( -\frac{\partial}{\partial \tau} - E \right)}{\det\left( -\frac{\partial}{\partial \tau}\right)} = \frac{\prod_{n = -\infty}^{\infty}\frac{(2n+1) i \pi}{\beta} - E}{\prod_{n = -\infty}^{\infty}\frac{(2n+1) i \pi}{\beta}}}\qquad\qquad \\ & = & \prod_{n = -\infty}^{\infty} \left( 1 + i \frac{\beta E}{(2n+1) \pi} \right) \\
  & = & \prod_{n = 0}^{\infty} \left[ 1 + \left( \frac{\beta E}{(2n+1) \pi} \right)^2  \right] \\
  & = & \frac{\prod_{n = 1}^{\infty} \left[ 1 + \left( \frac{\beta E}{n \pi} \right)^2  \right]}{\prod_{n = 1}^{\infty} \left[ 1 + \left( \frac{\beta E}{2 n \pi} \right)^2  \right]} \\
  & = & \frac{\sinh(\beta E)}{\sinh(\beta E / 2)} = 2 \cosh(\beta E / 2) \\
  & = & \exp(\beta E / 2) [ 1 + \exp(-\beta E)].
\end{eqnarray}

\section{Recursion relations}
\label{app: recursion}

In this Appendix we show that the quantity
$s_n(\left\{\left\{ E_{i_1}, E_{i_2}, \ldots, E_{i_n} \right\}\right\})$ given by Eq.\,(\ref{eq: matsum}) can be obtained recursively.
The recursion has the following desirable feature: at each step of recursion,
at most one singular denominator occurs, which allows to apply directly
l'H\^{o}pital's rule. The recursion provides an expression for $s_n$
for the multiset of energies $\epsilon = \left\{\left\{ E_{i_1}, E_{i_2}, \ldots, E_{i_n} \right\}\right\}$ in terms of $s_{n-1}$:
\begin{eqnarray}
  \lefteqn{s_n(\epsilon) = \frac{1}{n(n-1)} \sum_{a, b = 1, 2, \ldots, n \atop a \ne b} {}} \nonumber\\ & & \qquad \frac{s_{n-1}(\epsilon \setminus \{\{ E_{i_a} \}\}) - s_{n-1}(\epsilon \setminus \{\{ E_{i_b} \}\})}{E_{i_b} - E_{i_a}}.
\label{eq: recurs}
\end{eqnarray}
The summand has the form $\frac{g(E_{i_b}) - g(E_{i_a})}{E_{i_b} - E_{i_a}}$, with $g(y) = s_{n-1}(\epsilon \setminus \{\{ E_{i_a}, E_{i_b} \}\} \cup \{\{ y \}\})$.

Alternatively, the following representation in terms of the Vandermonde determinants is possible:
\begin{eqnarray}
  \lefteqn{s_n(\left\{\left\{ E_{1}, E_{2}, \ldots, E_{n} \right\}\right\}) = {}} \nonumber \\
  & &
    \left| \begin{array}{cccc} 1& 1& \ldots& 1\\ E_1& E_2& \ldots& E_n\\ (E_1)^2& (E_2)^2& \ldots& (E_n)^2\\
    \vdots& \vdots& \cdots& \vdots\\ (E_1)^{n-2}& (E_2)^{n-2}& \ldots& (E_n)^{n-2}\\
    f(E_1)& f(E_2)& \ldots& f(E_n) \end{array} \right|
  \Biggr/ \nonumber \\ & & \qquad {}
    \left| \begin{array}{cccc} 1& 1& \ldots& 1\\ E_1& E_2& \ldots& E_n\\ (E_1)^2& (E_2)^2& \ldots& (E_n)^2\\
    \vdots& \vdots& \cdots& \vdots\\ (E_1)^{n-1}& (E_2)^{n-1}& \ldots& (E_n)^{n-1} \end{array} \right|
  .
\end{eqnarray}
If a degeneracy is present in the set of energies,
a repeated column should be replaced by its consecutive derivatives with respect to the corresponding energy,
identically in the numerator and the denominator. For example, the expression (\ref{eq: reg xxyz}) can be written as
\begin{equation}
  \left.
    \left| \begin{array}{cccc} 1& 0& 1& 1\\ x& 1& y& z\\ x^2& 2 x& y^2& z^2\\
    f(x)& f'(x)& f(y)& f(z) \end{array} \right|
  \Biggm/
    \left| \begin{array}{cccc} 1& 0& 1& 1\\ x& 1& y& z\\ x^2& 2 x& y^2& z^2\\
    x^3& 3x^2& y^3& z^3 \end{array} \right|
  \right. ,
\end{equation}
and (\ref{eq: reg xxyy}) as
\begin{equation}
  \left.
    \left| \begin{array}{cccc} 1& 0& 1& 0\\ x& 1& y& 1\\ x^2& 2 x& y^2& 2 y\\
    f(x)& f'(x)& f(y)& f'(y) \end{array} \right|
  \Biggm/
    \left| \begin{array}{cccc} 1& 0& 1& 0\\ x& 1& y& 1\\ x^2& 2 x& y^2& 2 y\\
    x^3& 3x^2& y^3& 3 y^2 \end{array} \right|
  \right. .
\end{equation}

\section{Relation to Rayleigh-Schr\"{o}dinger perturbation theory}
\label{app: RS}

We will now demonstrate how the third order perturbation
can be derived using the standard Rayleigh-Schr\"{o}dinger (RS)  perturbation theory. The consecutive orders of expansion are,
\begin{widetext}
\begin{equation}
  E^{(1)}_{n} = \left< n \middle| V \middle| n \right>, \quad
  E^{(2)}_{n} = \sum_{k \ne n} \frac{\left| \left< k \middle| V \middle| n \right> \right|^2}{E_n - E_k},
\end{equation}
\begin{equation}
  E^{(3)}_n = \sum_{k \ne n} \sum_{m \ne n} \frac{\left< n \middle| V \middle| m \right>
  \left< m \middle| V \middle| k \right> \left< k \middle| V \middle| n \right>}{(E_n - E_m) (E_n - E_k)} -
  \left< n \middle| V \middle| n \right> \sum_{m \ne n} \frac{\left| \left< n \middle| V \middle| m \right> \right|^2}{(E_n - E_m)^2}.
\end{equation}
We see a subtracted term on the r.h.s. of the equation for $E^{(3)}_n$.
To expand $\Omega(\lambda)$ we need only the energies:
\begin{equation}
  \Omega(\lambda) = -\frac{1}{\beta} \sum_n \ln\left[ 1 + \exp( -\beta E_n(\lambda) ) \right].
\end{equation}
Indeed,
\begin{equation}
  \Omega^{(3)} = \sum_n \left[ f(E_n) E^{(3)}_n + f'(E_n) E^{(1)}_{n} E^{(2)}_{n} + \frac{1}{6} f''(E_n) \left(E^{(1)}_{n}\right)^3 \right].
\end{equation}
Direct substitution yields:
\begin{eqnarray} \lefteqn{
  \Omega^{(3)} = \left[ \sum_{k \ne n \atop m \ne n} \frac{f(E_n)}{(E_n - E_m) (E_n - E_k)} \left< n \middle| V \middle| m \right>
  \left< m \middle| V \middle| k \right> \left< k \middle| V \middle| n \right> + {} \right.} \nonumber \\
  & & {} \left. - \sum_{m \ne n} \frac{f(E_n)}{(E_n - E_m)^2}
   \left< n \middle| V \middle| n \right> \left| \left< n \middle| V \middle| m \right> \right|^2 + {} \right. \nonumber \\
   & & \left. {} + \sum_{m \ne n} \frac{f'(E_n)}{E_n - E_m} \left< n \middle| V \middle| n \right> \left| \left< m \middle| V \middle| n \right> \right|^2 + \sum_n \frac{1}{6} f''(E_n) \left(\left< n \middle| V \middle| n \right>\right)^3 \right].
\end{eqnarray}
We regroup the terms as follows:
\begin{eqnarray} \lefteqn{
  \Omega^{(3)} = \frac{1}{3} \left\{ \sum_{k \ne m \ne n} \left[ \frac{f(E_k)}{(E_k - E_m) (E_k - E_n)} + \frac{f(E_m)}{(E_m - E_k) (E_m - E_n)} + \frac{f(E_n)}{(E_n - E_k) (E_n - E_m)} \right] \times \right.} \qquad\qquad \nonumber \\
  & & {} \qquad \qquad \qquad \qquad \qquad \qquad \times \left< n \middle| V \middle| m \right>
  \left< m \middle| V \middle| k \right> \left< k \middle| V \middle| n \right> + {} \nonumber \\
  & & {} + \sum_{m \ne n} \left[ \frac{f(E_m) - f(E_n)}{(E_n - E_m)^2} + \frac{f'(E_n)}{E_n - E_m} \right]
   \left[ \left< n \middle| V \middle| n \right> \left< n \middle| V \middle| m \right> \left< m \middle| V \middle| n \right> + {} \right. \nonumber \\
  & & \qquad \qquad \qquad \qquad \qquad \left. {} + \left< m \middle| V \middle| n \right> \left< n \middle| V \middle| n \right>\left < n \middle| V \middle| m \right> + \left< n \middle| V \middle| m \right> \left< m \middle| V \middle| n \right>\left < n \middle| V \middle| n \right> \right] + {} \nonumber \\
   & & \left. {} + \sum_n \frac{f''(E_n)}{2} \left< n \middle| V \middle| n \right>\left< n \middle| V \middle| n \right>\left< n \middle| V \middle| n \right> \right\}.
\end{eqnarray}
\end{widetext}
This is the required form compatible with Eq.\,(\ref{eq: p3}). Indeed, the first term is the direct symmetrization of Eq.\,(\ref{eq: p3}). However, if $k$ or $m$ becomes equal $n$, one uses Eq.\,(\ref{eq: reg xxy}). Finally, if $k = m = n$, Eq.\,(\ref{eq: reg xxx}) applies. The purpose of repeating some terms is to make the combinatorial weights evident. We note that one can symmetrize either the part comprising the occupation function and the energetic denominators, or the product of matrix elements (or both).

\section{Relation to path integrals}
\label{app: path}

In this Appendix we relate our observation to the properties of path integrals.
The partition function $Z(\lambda)$ for a system of fermions can be written as a coherent state path integral,
\begin{equation}
  Z(\lambda) = \mathop{\mathrm{Tr}} e^{-\beta H} = \int \mathcal{D}[\xi_\alpha(\tau), \overline{\xi}_\alpha(\tau)]
    \, e^{\mathcal{S}[\xi_\alpha(\tau), \overline{\xi}_\alpha(\tau)]},
\end{equation}
where $\alpha$ labels fermionic (Grassmannian) degrees of freedom $\xi_\alpha$, $0 < \tau < \beta$
is the imaginary time, and $\mathcal{S}$ is the usual action
\begin{eqnarray}
   \lefteqn{\mathcal{S}[\xi_\alpha(\tau), \overline{\xi}_\alpha(\tau)]  {}}\nonumber \\ & & =\int_0^\beta d\tau \, \left( - \overline{\xi}_\alpha(\tau) \frac{\partial \xi_\alpha(\tau)}{\partial\tau} - H\left(\xi_\alpha(\tau), \overline{\xi}_\alpha(\tau)\right)\right). \qquad
\end{eqnarray}
In simple terms, this is to be understood as follows: the integration domain $[0, \beta]$ is divided into $N$ equal intervals,
each $\epsilon = \beta / N$ long. Then $\mathop{\mathrm{Tr}} e^{-\beta H} = \mathop{\mathrm{Tr}} \left[ e^{-\epsilon H} e^{-\epsilon H} \ldots e^{-\epsilon H} \right] = \mathop{\mathrm{Tr}} \left[ e^{-\epsilon H} \mathbbm{1} e^{-\epsilon H} \mathbbm{1} \ldots e^{-\epsilon H}
\mathbbm{1} \right]$, where $\mathbbm{1}$ is a resolution of unity, $\mathbbm{1} = \int d\mu(z) \left| z \right> \left< z \right|$,
and the trace turns out to be an $N$-dimensional integral (in the limit $N \to \infty$, a path integral).
Assume $H = H_0 + \lambda V$. We use the Trotter formula, $e^{-\epsilon H} \approx e^{-\epsilon H_0} e^{-\epsilon \lambda V}$, and approximate $e^{-\epsilon \lambda V} \approx 1 - \epsilon \lambda V$. This leads to
an integral representation for the terms of the expansion $Z(\lambda) = Z^{(0)} + \lambda Z^{(1)} + \lambda^2 Z^{(2)} + \ldots$ (from which an expansion of the free energy $F = -\frac{1}{\beta} \ln Z$ can be directly obtained through series composition):
\begin{eqnarray}
  Z^{(0)} & = & \mathop{\mathrm{Tr}} e^{-\beta H_0}; \\
  Z^{(1)} & = & \int_{\tau_1 = 0}^\beta d \tau_1\, \mathop{\mathrm{Tr}} \left[ e^{-\tau_1 H_0} (-V) e^{-(\beta - \tau_1) H_0} \right]; \\
  Z^{(2)} & = & \int_{\tau_1 = 0}^\beta d \tau_1 \int_{\tau_2 = \tau_1}^\beta d \tau_2 \mathop{\mathrm{Tr}} \Bigl[ e^{-\tau_1 H_0} (-V) e^{-(\tau_2 - \tau_1) H_0} {} \nonumber \\
  & & \qquad \times (-V) e^{-(\beta - \tau_2) H_0} \Bigr],
\end{eqnarray}
and $Z^{(3)}$ is given by a triple integral
\begin{eqnarray}
  \lefteqn{Z^{(3)} = \int_{\tau_1 = 0}^\beta d \tau_1 \int_{\tau_2 = \tau_1}^\beta d \tau_2 \int_{\tau_3 = \tau_2}^\beta d \tau_3 \, \mathop{\mathrm{Tr}} \left[ e^{-\tau_1 H_0} (-V) {} \right.} \nonumber  \\
   & & \times \left. e^{-(\tau_2 - \tau_1) H_0} (-V) e^{-(\tau_3 - \tau_2) H_0} (-V) e^{-(\beta - \tau_3) H_0} \right]. \qquad \label{eq: zztord}
\end{eqnarray}
At this point we can substitute a diagonal non-interacting $H_0$ and a one-particle fermionic operator $V$ into these expressions. We obtain (using a computer algebra system), in the generic case $k \ne m \ne n$, terms in $\Omega^{(3)}$ of the form
\begin{eqnarray}
  \lefteqn{\left[ \frac{f(E_k)}{(E_k - E_m) (E_k - E_n)} + \frac{f(E_m)}{(E_m - E_k) (E_m - E_n)} + {} \right.} \nonumber
  \\ & & \left. {} + \frac{f(E_n)}{(E_n - E_k) (E_n - E_m)} \right]  {} \nonumber \\
  & & \qquad\qquad\qquad\qquad \times\left< k \middle| V \middle| m \right> \left< m \middle| V \middle| n \right> \left< n \middle| V \middle| k \right>,
  \quad \label{eq: 3ok}
\end{eqnarray}
rather than
\begin{eqnarray}
\lefteqn{\left\{ \frac{f(E_k) [1 - f(E_m)] [1 - f(E_n)]}{(E_k - E_m) (E_k - E_n)} + c.p. \right\} {} \times}  \nonumber\\ & & \qquad \qquad {} \times
\left< k \middle| V \middle| m \right> \left< m \middle| V \middle| n \right> \left< n \middle| V \middle| k \right>. \label{eq: 3bad}
\end{eqnarray}
Accidentally, Eq.\,(\ref{eq: 3ok}) is equivalent to the particle-hole symmetrized form of Eq.\,(\ref{eq: 3bad}) obtained by taking a difference of Eq.\,(\ref{eq: 3bad})  and the same expression with negated energies (this observation is specific to the third order perturbation).

\bibliography{my4th_br08Feb2018}

\end{document}